\documentclass[aps,pra,showpacs,twocolumn]{revtex4}
\AtBeginDvi{}
\usepackage[dvips]{graphicx}
\usepackage{dcolumn}
\usepackage{bm}
\usepackage{amsmath,amssymb}
\usepackage{enumerate}
\usepackage{booktabs}
\usepackage{epsfig}
\begin{document}
\title{\textit{p}-wave superfluid and phase separation in atomic Bose-Fermi mixture}
\author{Kazunori Suzuki$^1$, Takahiko Miyakawa$^2$, and Toru Suzuki$^1$}
\affiliation{
$^1$Department of Physics, Tokyo Metropolitan University, 1-1 Minami-Osawa, Hachioji, Tokyo 192-0397, Japan
\\
$^2$Department of Physics, Faculty of Science, Tokyo University of Science, 1-3 Kagurazaka, Shinjuku, Tokyo 162-8601, Japan}
\date{\today}
\begin{abstract}
We consider a system of repulsively interacting Bose-Fermi mixtures of spin polarized uniform
atomic gases at zero temperature. 
We examine possible realization of $p$-wave superfluidity of fermions due to an effective 
attractive interaction via density fluctuations of Bose-Einstein condensate within mean-field 
approximation. We find the ground state of the system by direct energy comparison 
of $p$-wave superfluid and phase-separated states, and suggest an occurrence of the $p$-wave 
superfluid for a strong boson-fermion interaction regime. We study some signatures in 
the $p$-wave superfluid phase, such as anisotropic energy gap and quasi-particle energy in 
the axial state, that have not been observed in spin unpolarized superfluid of atomic fermions.
We also show that a Cooper pair is a tightly bound state like a diatomic molecule in the strong boson-fermion coupling regime and suggest an observable indication of the $p$-wave superfluid in the real experiment.
\end{abstract}
\maketitle
%
%
\section{Introduction}
Magnetically tunable Feshbach resonances have opened up a new field of research in the 
physics of ultracold atomic gases that exhibits exciting  phenomena. 
In two-component fermi gases, for instance,  the condensate of atom pairs in the BCS-BEC 
crossover regime has been intensively studied~\cite{BCS-BECex1,BCS-BECex2,BCS-BECex3,BCS-BECth1,BCS-BECth2,BCS-BECth3}. 
In an ultracold atomic system, the \textit{p}-wave Feshbach resonance was also found, 
which allows one to tune \textit{p}-wave interactions between atoms in the spin-polarized 
fermi system~\cite{P-Feshbach}. It has now become possible to produce $p$-wave molecules 
between $^{40}$K atoms using this technique~\cite{P-molecule}.

Atomic boson-fermion mixed systems, too, are expected to show many interesting phenomena.
In this mixed system, it has been reported the observation of simultaneous quantum degeneracy 
of Bose-Einstein condensate (BEC) and Degenerate Fermi gas~\cite{BEC-FD,Hadzibabic,BEC-FD-He}, and the 
collapse of fermions in the attractively interacting mixture~\cite{Instability1,Instability2}.
On the theoretical side a number of studies on the static and dynamical properties of the mixture 
have been made~\cite{Molmer,Amoruso,Miyakawa,Roth,Sogo}. We note here that the strongly coupled boson-fermion 
pair may behave as a heteronuclear molecule or resonance, and its role in the mixture has been 
studied~\cite{Yabu,BF-FFLO,Kondo,BFpair1,BFpair2}. Recent experiments show an existence of the 
Feshbach resonances between bosons and fermions~\cite{BF-Feshbach-NaLi,BF-Feshbach-RbK}, 
and quite recently a formation of the boson-fermion heteronuclear molecule in the optical lattice 
has been reported~\cite{BFmolecule}. As in the case of fermionic system, the existence of the 
Feshbach resonance allows one to control the boson-fermion interaction~\cite{RbK-tune,RbK-cont}.
Tuning the interaction one may induce a collapse of fermions or a phase separation of
bosons and fermions, depending on the sign of the interaction~\cite{Molmer,Miyakawa,Roth,Binary-mixture}.

In many experiments of atomic Bose-Fermi mixture, all atomic fermions have the same spin 
components as they are trapped by magneto-optical trap together with bosons. The fermion-fermion 
interaction has been negligible since the \textit{s}-wave scattering amplitude between fermions 
vanishes because of Pauli exclusion principle, and the other higher partial waves do not contribute at low 
temperatures. In a boson-fermion mixed system, boson density fluctuations may give rise to an attractive 
interaction between fermions as in the electron-phonon system, and may induce a 
superfluid transition~\cite{Binary-mixture,FBF-pair}. This mechanism has been 
studied in the liquid $^3 \rm He$-$^4 \rm He$ mixtures~\cite{3He-4He}. If this mechanism is 
strong enough in the  ultracold atomic mixture, \textit{p}-wave Cooper pairs will be 
formed between spin-polarized fermions~\cite{pFBF-pair,pFBF-pair2,pFBF-pair3}. We should note, 
however, that a realization of the superfluid depends on a balance of different mechanisms
as mentioned above.
A sufficiently strong attractive boson-fermion interaction, for instance, may cause a collapse
of the system rather than the superfluid state. On the other hand, a strong boson-fermion 
repulsion may favor a phase separation of bosons and fermions instead of the superfluid transition.

The purpose of the present paper is to investigate a possibility of \textit{p}-wave superfluid 
transition in the repulsively interacting spin-polarized Bose-Fermi mixture at zero temperature.
We consider superfluid transition induced by a Bogoliubov phonon-mediated fermion-fermion interaction, 
and compare energies of the $p$-wave superfluid state and the boson-fermion phase-separated state. 
In view of the recent development in the tuning of boson-fermion interaction via Feshbach 
resonances, we study the system in a broad range of the interaction parameters. 
Earlier study in this direction shows that the \textit{p}-wave superfluid transition is 
hard to occur, i.e., its transition temperature is too low to be attainable, because 
the phonon-mediated attractive interaction is very weak~\cite{pFBF-pair}. A stronger 
repulsive interaction would cause an instability of the system towards phase separation, 
and then the induced attractive interaction reduces. We show, however, that for a very strong 
repulsive interaction it is possible to realize the \textit{p}-wave superfluid transition.
In the superfluid phase, anisotropic energy gap appears like that of \textit{p}-wave superfluid
of $^3$He and heavy-fermion systems.
We also show that the Cooper pair is a tightly bound state like a diatomic molecule in the strong boson-fermion coupling regime~\cite{Leggett,Noz-Schmi}.

Main content of the present paper is as follows.
Sec.II derives ground state energies of the system in the $p$-wave superfluid phase due to
the phonon-mediated attractive interaction and in the normal phase by the method of Ref.\cite{Binary-mixture}.
Sec.III presents phase diagram of the system by direct energy comparison of the $p$-wave superfluid and
phase-separated states. 
The momentum dependence of the energy gap and other properties of the superfluid state are shown.
Finally, Sec. IV is a summary.
%
%
%
%
\section{Model}
We consider a spin-polarized uniform mixture of atomic bosons of mass $m_{\rm B}$ and fermions of mass $m_{\rm F}$ at zero temperature.
The system is described by the Hamiltonian
\begin{equation}
      \hat H = \hat H_{\rm F} + \hat H_{\rm B} + \hat H_{int},
\end{equation}
where
\begin{align}
      \hat H_{\rm F}
      &= \sum_{\mathbf k} ( \epsilon_{\mathbf k}^f - \mu_{\rm F} ) c_{\mathbf k}^\dagger c_{\mathbf k}^{},
\\
      \hat H_{\rm B}
      &= \sum_{\mathbf k} ( \epsilon_{\mathbf k}^b - \mu_{\rm B} ) b_{\mathbf k}^\dagger b_{\mathbf k}^{}
      + \frac{ U_{\rm BB} }{ 2 } \sum_{\mathbf k \mathbf p \mathbf q}
                          b_{ \mathbf{p+q} }^\dagger b_{ \mathbf{k-q} }^\dagger
                          b_{ \mathbf p }^{} b_{ \mathbf k }^{},
\\
      \hat H_{int} &= U_{\rm BF} \sum_{\mathbf k \mathbf p \mathbf q}
                          b_{ \mathbf{p+q} }^\dagger c_{ \mathbf{k-q} }^\dagger
                          b_{ \mathbf p }^{} c_{ \mathbf k }^{}.
\end{align}
Here $c_{\mathbf k}^{(\dagger)}$ and $b_{\mathbf k}^{(\dagger)}$ are the annihilation (creation) operators for
the fermionic and bosonic atoms of momentum $\bf k$, respectively. 
The corresponding momentum states have kinetic energies
of bosons $\epsilon^b_{\bf k}=\hbar^2 {\bf k}^2/2m_{\rm B}$
and of fermions $\epsilon^f_{\bf k}=\hbar^2 {\bf k}^2/2m_{\rm F}$.
The chemical potentials of fermions and bosons are denoted by $\mu_{\rm F}$ and $\mu_{\rm B}$.
The  boson-boson and boson-fermion collisions are described by the interaction strengths
$U_{\rm BB}=4\pi\hbar^2 a_{\rm BB}/m_{\rm B}$ and $U_{\rm BF}=2\pi\hbar^2 a_{\rm BF}/m_r$, respectively,
where $a_{\rm BB}$ and $a_{\rm BF}$ are the corresponding $s$-wave scattering lengths and $m_r=m_{\rm B}m_{\rm F}/(m_{\rm B}+m_{\rm F})$ is the reduced mass.
The elastic fermion-fermion $s$-wave scattering for spin-polarized fermions is absent
because of Pauli exclusion principle.

In order to give a realistic estimate of the phase diagram, we consider a system of $ \rm ^{87}Rb $-$ \rm ^{40}K $
mixture for the atomic bosons and fermions, where $ m_{\rm B} = 1.419 \times 10^{-25}~ \rm kg $ and $ m_{\rm F}= 0.649 \times 10^{-25}~ \rm kg $, and $a_{\rm BB} = 98.98 a_0 $ with Bohr radius $a_0$.
As for the boson-fermion interaction, we assume the scattering length to be tunable via Feshbach resonance.
This can be realized in the current experimental situation~\cite{BF-Feshbach-NaLi,BF-Feshbach-RbK,RbK-tune,RbK-cont}.
Hereafter, we take $a_{\rm BF}>0$, leading to repulsively interacting Bose-Fermi mixtures.

We consider two types of quantum phases of fermions, that is, the normal phase and the superfluid phase.
On the one hand, a phase separation of bosons and fermions occurs in the normal state of fermions,
as discussed in Ref.~\cite{Binary-mixture} for a large positive $U_{\rm BF}$.
On the other hand, for a strongly repulsive boson-fermion coupling,
an effective fermion-fermion interaction via density fluctuations of BEC
becomes strong and is expected to lead to a superfluid state of fermions.
In the following we calculate ground state energies of these two types of phases, separately~\cite{note1}.
\subsection{Superfluid phase : $p$-wave pairing}
We assume that all bosons are condensate and the fluctuations are treated by excitations of
Bogoliubov phonon of energy: $\hbar \omega_{\mathbf k}^b = \sqrt{ \epsilon_{\mathbf k}^b
 ( \epsilon_{\mathbf k}^b + 2 U_{\rm BB} n_{\rm B} ) }$ where $n_{\rm B}$ is a condensate density.
Thus the bosonic part of the Hamiltonian, Eq.(3), is transformed to 
\[
\hat H_{\rm B} =
      - \mu_{\rm B} n_{\rm B} + U_{\rm BB} n_{\rm B}^2 / 2
      + \sum_{\mathbf k \neq 0} \hbar \omega_{\mathbf k}^b \beta_{\mathbf k}^\dagger \beta_{\mathbf k}^{},
\]
and the boson-fermion interaction Hamiltonian, Eq.(4), is also transformed to
\begin{align}
\label{EffectiveBF}
      \hat H_{int}
      &\simeq
      U_{\rm BF} n_{\rm B} \sum_{\mathbf k} c_{\mathbf k}^\dagger c_{\mathbf k}^{}
\notag
\\
      &+ U_{\rm BF} \sqrt n_{\rm B} \sum_{\mathbf{ k \neq 0, p } }
        \sqrt{ \frac{ \epsilon_{\mathbf k}^b }{ \hbar \omega_{\mathbf k}^b } }
        \left( \beta_{\mathbf k}^{} c_{ \mathbf{k+p} }^\dagger c_{\mathbf p}^{}
        + \beta_{\mathbf k}^\dagger c_{ \mathbf p }^\dagger c_{ \mathbf{k+p} }^{} \right),
\end{align}
where $\beta_{\mathbf k}^{(\dagger)}$ is a Bogoliubov phonon annihilation (creation) operator.
In these transformations, we have neglected higher order terms of $\beta_{\bf k}^{(\dagger)}$.
We regard the first term of Eq.(\ref{EffectiveBF}) as an energy shift of 
the fermionic chemical potential $\mu_{\rm F}^r \equiv \mu_{\rm F} - U_{\rm BF} n_{\rm B}$.

By applying the second-order perturbation theory, the phonon-mediated interaction between 
two fermions is derived as one-phonon exchange process,
\[
      \hat H_{int} =
      - \frac{ 1 }{ 2 } \sum_{ \mathbf p \mathbf p^\prime \mathbf q } U_{\rm FBF}( \mathbf{p^\prime, q} )
                           c_{ \mathbf{p+q} }^\dagger c_{ \mathbf{p^\prime-q} }^\dagger
                           c_{ \mathbf{p^\prime} }^{} c_{ \mathbf p }^{},
\]
with
\[
      U_{\rm FBF}(\mathbf{p^\prime, q}) =
      U_{\rm BF}^2 n_{\rm B} \frac{ 2 \epsilon_{\mathbf q}^b }
 { (\hbar \omega_{\mathbf q}^b)^2 - ( \epsilon_{ \mathbf{p^\prime-q} }^f - \epsilon_{ \mathbf{p^\prime} }^f )^2 }.
\]
We assume that Fermi velocity $v_{\rm F}=\hbar k_{\rm F} /m_{\rm F}$ is much smaller
than Bogoliubov sound velocity $s=\sqrt{U_{\rm BB} n_{\rm B} / m_{\rm B}}$ , i.e. $v_{\rm F}  \ll s$,
so the phonon-mediated effective interaction can be written as in Ref.\cite{pFBF-pair},
\begin{equation}
\label{Yukawatype}
      U_{\rm FBF}(\bm q)
              = \frac{ U_{\rm BF}^2 }{ U_{\rm BB} } \frac{ 1 }{ 1 + \left( \hbar |{\bm q}| / 2 m_{\rm B} s \right)^2}.
\end{equation}
This interaction is equivalent to an effective interaction between fermions induced by density fluctuations of
background of BEC when the retardation effect is neglected~\cite{Binary-mixture,FBF-pair}.
The effective Hamiltonian of fermions is then given by
\begin{align}
\label{effectiveH}
      &\hat H_{\rm F}^{eff}
          = \sum_{\mathbf k} ( \epsilon_{\mathbf k}^f - \mu_{\rm F}^r ) c_{\mathbf k}^\dagger c_{\mathbf k}^{}
\notag
\\
            &- \frac{ 1 }{ 2 } \sum_{\mathbf P \mathbf k \mathbf k^\prime} U_{\rm FBF}( \mathbf{ k-k^\prime } )
                          c_{ \mathbf{P/2+k} }^\dagger c_{ \mathbf{P/2-k} }^\dagger
                          c_{ \mathbf{P/2-k^\prime} }^{} c_{ \mathbf{P/2+k^\prime} }^{}.
\end{align}

Since we consider spin-polarized fermions, the effective interaction in the channel with even angular
momentum $l$ is absent due to antisymmetrization of the orbital wave function in the relative coordinate.
Furthermore for the effective interaction described by Eq.~(\ref{Yukawatype}) the contribution to
the interaction for higher $l$ can be negligible~\cite{FBF-pair}.
Thus we extract the dominant $l=1$ component of the phonon-mediated interaction
\begin{equation}
\label{pwaveinteraction}
      U_{\rm FBF}^{\textit p}( \mathbf{ k,k^\prime } )
      = 3 U_{ind}^{\textit p} (k,k^\prime)
        \sum_{i=x,y,z} \hat k_i^{} \hat k_i^\prime
\end{equation}
with
\begin{align}
      &U_{ind}^{\textit p} (k,k^\prime)
      = \frac{ U_{\rm BF}^2 }{ U_{\rm BB} } \frac{ 2 m_{\rm B}^2 s^2 }{ \hbar^2 k k^\prime }  \times
\notag
\\
    &        \left( \frac{ k^2 + k^{\prime 2} + ( 2 m_{\rm B} s /\hbar )^2 }{ 4 k k^\prime }
                     \ln \left| \frac{ ( k + k^\prime )^2 + ( 2 m_{\rm B} s /\hbar )^2 }
                                     { ( k - k^\prime )^2 + ( 2 m_{\rm B} s /\hbar )^2 } \right| - 1 \right),\notag
\end{align}
where $k=|\bf k|$ and $\hat k_i=(\hat{\bf k} / |{\bf k}|)_i$.
%
In this approximation, only $p$-wave pairing is possible to realize.

Let us consider a Cooper pair with zero center-of-mass momentum and introduce a $p$-wave pair energy gap as
\[
\Delta(\mathbf k) = \sum_{\mathbf k^\prime} U_{\rm FBF}^{\textit p}( \mathbf{ k,k^\prime } ) \langle c_{ \mathbf{-k}^\prime }^{} c_{ \mathbf k^\prime }^{} \rangle
\]
where $\langle \ \rangle$ denotes an expectation value in the ground state.
In the standard BCS theory the effective Hamiltonian~Eq.(\ref{effectiveH}) can be diagonalized by Bogoliubov transformation 
$ \alpha_{\mathbf k} = u_{\mathbf k} c_{\mathbf k}^{} - v_{\mathbf k} c_{\mathbf{-k}}^\dagger $,
\begin{equation}
      \hat H_{\rm F}^{eff}
      = \sum_{\mathbf k}
      E_{\mathbf k} \alpha_{ \mathbf k }^\dagger \alpha_{ \mathbf k }^{}
    + \frac{ 1 }{ 2 } \sum_{\mathbf k} \left( \xi_{\mathbf k} - E_{\mathbf k}
    + \frac{ |\Delta({\mathbf k})|^2 }{ 2 E_{\mathbf k} } \right),
\end{equation}
where the quasi-particle energy is
$ E_{\mathbf k} = \sqrt{ \xi_{\mathbf k}^2 + |\Delta(\mathbf k)|^2 } $ 
with $ \xi_{\mathbf k} = \epsilon_{\mathbf k}^f - \mu_{\rm F}^r $, 
and $u_{\mathbf k}$ and $v_{\mathbf k}$ are given by
$ u_{\mathbf k} = \sqrt{ ( 1 + \xi_{\mathbf k} / E_{\mathbf k} )/2 } $ 
and
$ v_{\mathbf k} = \sqrt{ ( 1 - \xi_{\mathbf k} / E_{\mathbf k} )/2 } $,
respectively.
The gap equation
\begin{equation}
\label{gapequation}
      \Delta(\mathbf k) = \sum_{\mathbf k^\prime} U_{\rm FBF}^{\textit p}( \mathbf{ k,k^\prime } )
                          \frac{ \Delta(\mathbf k^\prime) }{ 2 E_{\mathbf k^\prime} },
\end{equation}
and the requirement of the mean number of fermions
\begin{equation}
\label{Nconservation}
      n_{\rm F} = \frac{1}{2} \sum_{\mathbf k} \left( 1 - \frac{ \xi_{\mathbf k} }
                                                               { E_{\mathbf k} } \right),
\end{equation}
determine the energy gap $\Delta(\mathbf k)$ and chemical potential of fermions $\mu_{\rm F}^r$.
Concrete forms of the gap will be discussed in the next section.

Once we have the optimized value of the energy gap and chemical potential,
we obtain the ground state energy per volume in the superfluid phase as
\begin{eqnarray}
\label{superfluidE}
      E_{SF}  &=& \frac{ 1 }{ 2 } \sum_{\mathbf k} \left( \xi_{\mathbf k} - E_{\mathbf k}
           + \frac{ |\Delta(\mathbf k)|^2 }{ 2 E_{\mathbf k} } \right) \nonumber \\
       &+& \mu_{\rm F}^r n_{\rm F}
                           + \frac{1}{2} U_{\rm BB} n_{\rm B}^2 + U_{\rm BF} n_{\rm B} n_{\rm F}.
\end{eqnarray}
The third and fourth terms in Eq.~(\ref{superfluidE}) are mean-field energies
of the boson-boson and boson-fermion interactions.
\subsection{Normal phase :  phase separation}

Now we consider a normal phase of repulsively interacting Bose-Fermi mixtures.
In Ref.~\cite{Binary-mixture}, Viverit~{\it et~al.} studied a phase diagram of the uniform mixtures at zero temperature and predicted three types of equilibrium states:
(A) a single uniform mixed phase, (B) a purely fermionic phase coexisting with a mixed phase,
and (C) a purely fermionic phase coexisting with a purely bosonic one.

By the method to find equilibrium states in Ref.~\cite{Binary-mixture},
we obtain the ground state energy of the system of a finite volume $V$
and total number of bosons $N_{\rm B}$ and of fermions $N_{\rm F}$.
Suppose that the system is composed of two phases of volume $V_1$ and $V_2$ where $V=V_1+V_2$.
The numbers of bosons and fermions are given by $N_{{\rm B}i}$ and $N_{{\rm F}i}$ for $i{\rm \text{-}th}$ phase, respectively, 
yielding to $N_{\rm B}=N_{\rm B1}+N_{\rm B2}$ and $N_{\rm F}=N_{\rm F1}+N_{\rm F2}$.
Thus the total densities of bosons $n_{\rm B} = N_{\rm B}/V$ and of fermions $n_{\rm F} = N_{\rm F}/V$ 
are given by $n_{\rm B} = n_{\rm B1} v + n_{\rm B2} (1-v)$ and  $n_{\rm F} = n_{\rm F1} v + n_{\rm F2} (1-v)$,
respectively, where $v = V_1/V$.
Total energy per volume~$V$ is given by
\begin{equation}
\label{normalenergy1}
E_N = E_1 v + E_2 (1-v),
\end{equation}
where the energy per volume~$V_i$ of   $i{\rm \text{-}th}$ phase has
\begin{equation}
\label{normalenergy2}
      E_i = \frac{ 3 }{ 5 } \epsilon_{{\rm F}i} n_{{\rm F}i}
          + \frac{ 1 }{ 2 } U_{\rm BB} n_{{\rm B}i}^2 + U_{\rm BF} n_{{\rm B}i} n_{{\rm F}i}.
\end{equation}
In Eq.~(\ref{normalenergy2}), the first term is the kinetic energy of fermions where
$\epsilon_{{\rm F}i}=\hbar^2 (6\pi^2 n_{{\rm F}i})^{2/3} / 2 m_{\rm F}$ is Fermi energy of  $i{\rm \text{-}th}$ phase.
The second and third terms correspond to the mean-field energies of the boson-boson
and boson-fermion interactions.
In  $i{\rm \text{-}th}$ phase, the pressure is $P_i = - \partial (E_i V_i) / \partial V_i$ and the chemical potentials
of bosons and fermions are given by $\mu_{{\rm B}i} = \partial E_i / \partial n_{{\rm B}i}$
and $\mu_{{\rm F}i} = \epsilon_{{\rm F} i}= \partial E_i / \partial n_{{\rm F}i}$, respectively.
The equilibrium conditions of the pressure and chemical potentials of bosons and fermions between two phases,
$P_1=P_2$, $\mu_{{\rm B}1}= \mu_{{\rm B}2} $, and $\epsilon_{{\rm F}1}= \epsilon_{{\rm F}2}$,
determine $V_i$, $n_{{\rm B} i}$, and $n_{{\rm F} i}$.
As a result, the three types of equilibrium states can be realized.

\begin{figure}
\begin{center}
\begin{flushleft}
\quad(a)
\end{flushleft}
\includegraphics[width=8.cm]{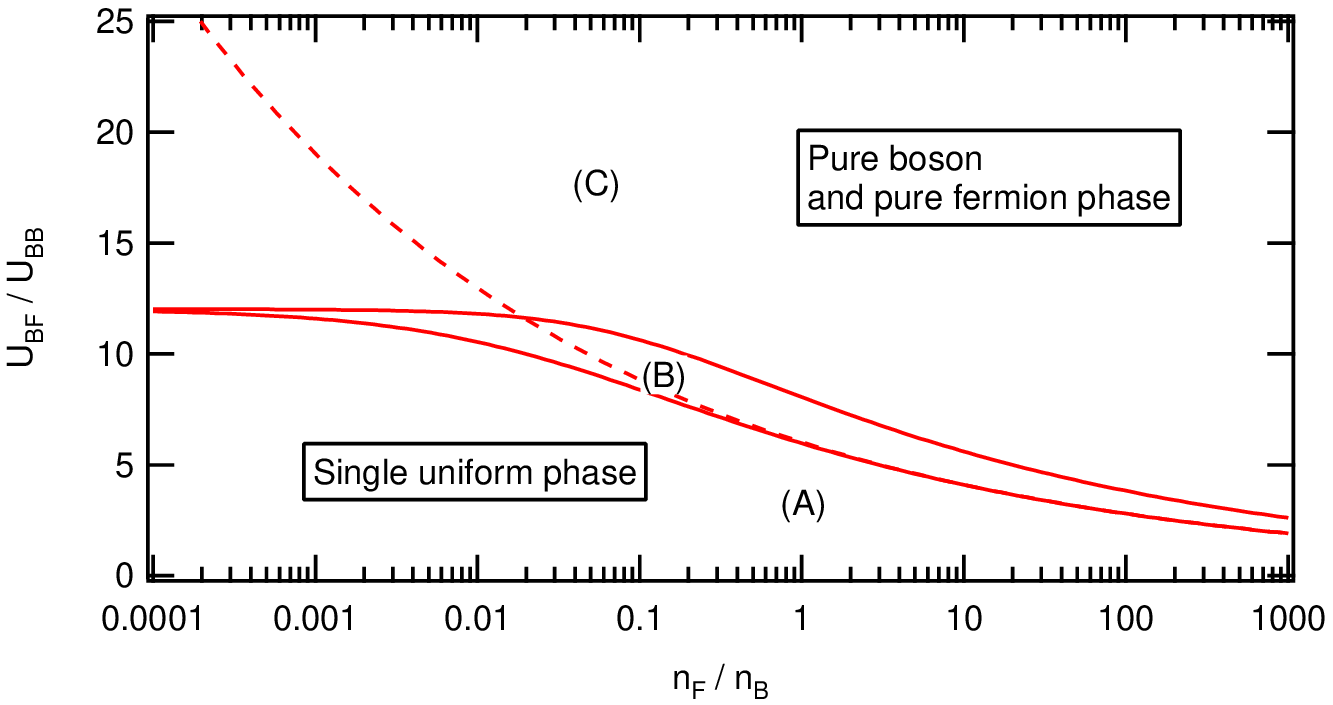}
\begin{flushleft}
\quad(b)
\end{flushleft}
\includegraphics[width=8.cm]{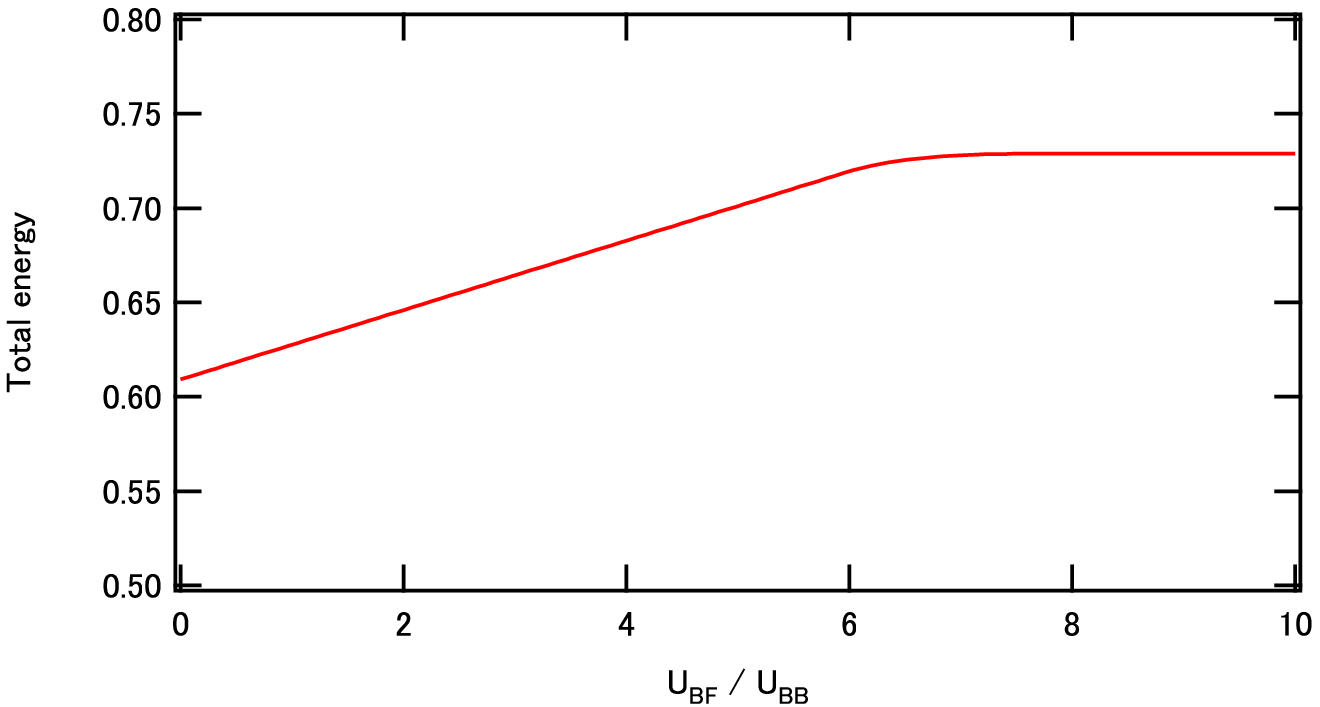}
\end{center}
\caption{
(a) Phase diagram of normal phase of the $^{87}$Rb-$^{40}$K mixtures.
There exist three types of equilibrium states: (A) a~single uniform mixed phase,
(B) a~purely fermionic phase coexisting with a mixed phase,
and (C) a~purely fermionic phase coexisting with a purely bosonic one.
The lower and upper solid lines in Fig.~1(a) represent the boundaries between the phases (A)-(B)
and phases (B)-(C), respectively.
The dashed line corresponds to a critical curve above which the uniform mixture is unstable against
small density fluctuations. 
(b) Total energy per volume of the ground state $E_N$ scaled by $\epsilon_{\rm F} n_{\rm F}$ as a function of $U_{\rm BF}/U_{\rm BB}$ for
$ n_{\rm F} = n_{\rm B} = 10^{14}  {\rm cm^{-3}}$.
}
\label{fig1}
\end{figure}

Figure~\ref{fig1}(a) shows a phase diagram for the  $^{87}\rm Rb$-$^{40}\rm K$ mixed system
in the $n_{\rm F}/n_{\rm B}$ vs. $U_{\rm BF}/U_{\rm BB}$ space for $n_{\rm B} = 10^{14} \rm cm^{-3}$
and $U_{\rm BB}=5.157 \times 10^{-51}$ [$\rm Jm^3$].
The lower and upper solid lines in Fig.~\ref{fig1}(a) represent the boundaries between the phases (A)-(B)
and phases (B)-(C), respectively.
The result shows that a phase separation of bosons and fermions is preferred for a higher fraction of fermions
and for a stronger boson-fermion interaction strength compared to the boson-boson one.
The dashed line corresponds to the critical curve above which the uniform mixture is unstable against
small density fluctuations.
%

%
%
Figure~\ref{fig1}(b) shows total energy per volume of the ground state $E_N$ calculated from 
Eqs.~(\ref{normalenergy1}) and (\ref{normalenergy2}) as a function of $U_{\rm BF}/U_{\rm BB}$
for $ n_{\rm F} = n_{\rm B} = 10^{14}$ $\rm cm^{-3}$.
The total energy per volume of the ground state scaled by $\epsilon_{\rm F} n_{\rm F}$ with $\epsilon_{\rm F} = \hbar^2 (6\pi^2 n_{\rm F})^{2/3} / 2 m_{\rm F}$ is plotted against the interaction ratio $U_{\rm BF}/U_{\rm BB}$.
The energy value increases as the repulsive boson-fermion interaction becomes stronger
until $U_{\rm BF}/U_{\rm BB} \simeq 6$ above which fermions start to separate from the bosonic cloud.
For $U_{\rm BF}/U_{\rm BB} \gtrsim 8$ the ground state is the completely phase-separated phase (C), and the energy becomes independent of the interaction ratio $U_{\rm BF}/U_{\rm BB}$.
%
%
%
%
%
\section{Results}

In this section, we examine phase transition between normal state and superfluid state of fermions.
To do so, we assume two typical types of energy gaps, an axial state and a polar state in the superfluid phase,
and calculate ground state energies in the two phases and directly compare them.
We obtain phase diagram of $^{87}$Rb-$^{40}$K mixtures showing that phase transition from
phase-separated state in the normal phase to $p$-wave superfluid state occurs
as repulsive boson-fermion coupling becomes stronger.
The unique characters of the $p$-wave pairing,
such as momentum dependence of energy gap, momentum distribution of fermions,
and quasi-particle excitation spectrum, in a strong boson-fermion coupling regime are also discussed.

%
\subsection{Transition from phase separation to $p$-wave superfluidity}

In order to obtain ground state energy of superfluid fermions, we need to calculate the total
energy described by Eq.~(\ref{superfluidE}).
In general, the $p$-wave pair energy gap can be expanded by spherical harmonics
\[
   \Delta(\mathbf k) = \sum_{m=-1}^{1} \Delta^m (k) Y_{1,m}(\theta,\phi).
 \]
We assume, however, just two typical types of pair energy gaps, so-called an axial state~\cite{ABM}
\begin{equation}
   \Delta^{\rm ax}(\mathbf k) = \Delta^{\rm ax}(k) ~( \hat k_x + i \hat k_y ),
\end{equation}
and a polar state
\begin{equation}
    \Delta^{\rm pl}(\mathbf k) = \Delta^{\rm pl}(k) ~\hat k_z,
\end{equation}
and regard a lower energy state of them as the ground state in superfluid phase.
%
%
Note that there is no symmetrical solution due to the spin-polarization of fermions in this system.
This fact is in stark contrast to unpolarized Fermi systems such as $^3 \rm He$ and
heavy-fermion systems~\cite{BW}.
Amplitudes of the axial and polar states have $|\Delta^{\rm ax}(\mathbf k)| = \Delta^{\rm ax}(k) ~|\sin \theta|$ and $|\Delta^{\rm pl}(\mathbf k)| = \Delta^{\rm pl}(k) ~|\cos \theta|$,
corresponding to anisotropic energy gaps in momentum space with zero energy gap
on the north and south poles for the axial state and on the equator for the polar state, respectively.

\begin{figure}
\begin{center}
\begin{flushleft}
\quad(a)
\end{flushleft}
\includegraphics[width=9.cm]{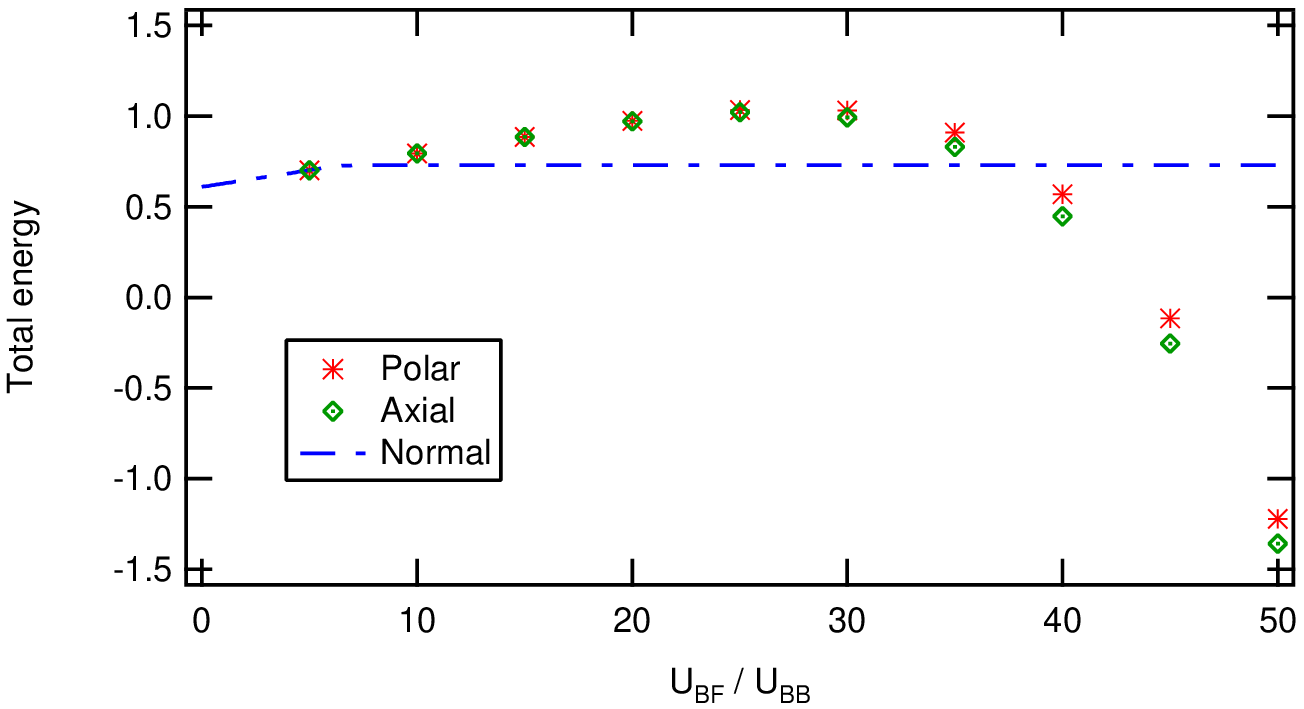}
\begin{flushleft}
\quad(b)
\end{flushleft}
\includegraphics[width=9.cm]{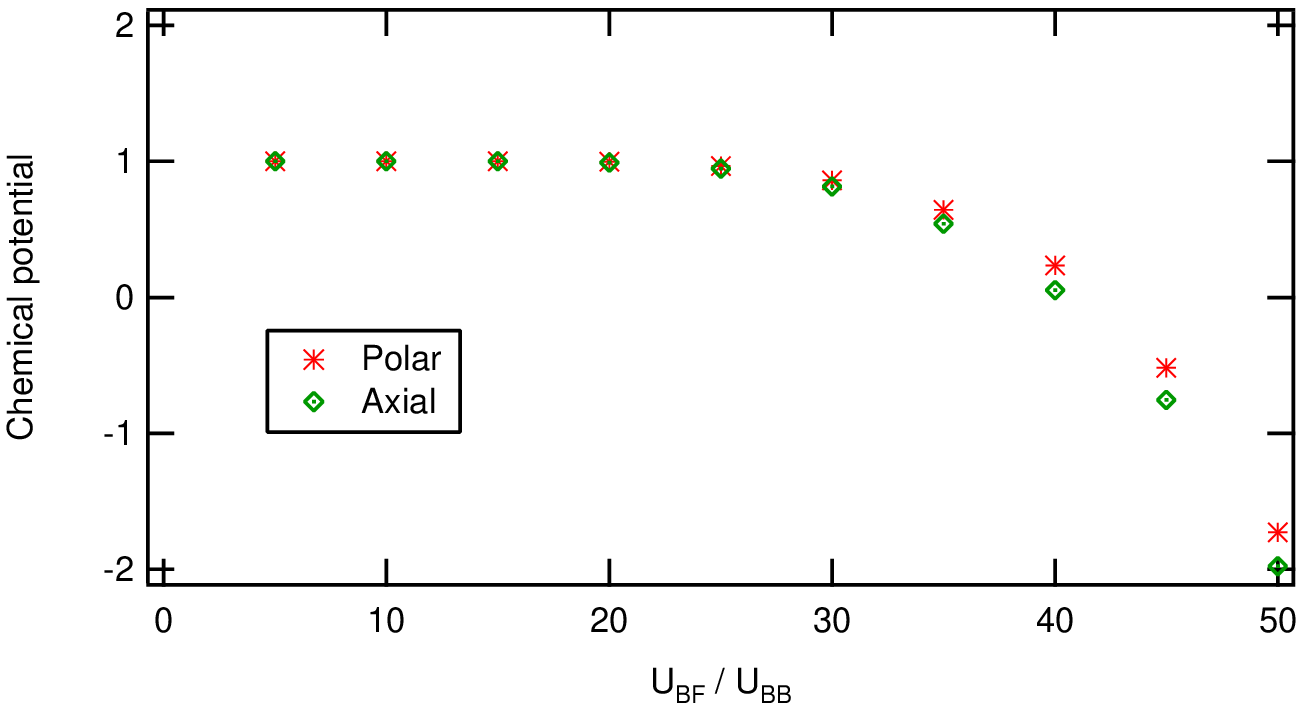}
\end{center}
\caption{
(a) Total energy per volume of the system $E_{SF}$ scaled by $\epsilon_{\rm F} n_{\rm F}$ and (b) chemical potential $\mu_{\rm F}^r$ scaled by $\epsilon_{\rm F}$ as a function of $U_{\rm BF}/U_{\rm BB}$ at $n_{\rm F} = n_{\rm B} = 10^{14} \rm cm^{-3}$.
Each mark of a tetragon and an asterisk denotes the axial state and polar one.
}
\label{fig2}
\end{figure}

Figure~\ref{fig2}(a) shows total energies per volume of the axial state (tetragon) and polar state (asterisk) in unit of $\epsilon_{\rm F} n_{\rm F}$ as a function of $U_{\rm BF}/U_{\rm BB}$ for $n_{\rm F}=n_{\rm B}=10^{14} {\rm cm^{-3}}$.
It shows that the total energies per volume of both of axial and polar states increase as the boson-fermion coupling becomes stronger until $U_{\rm BF}/U_{\rm BB} \simeq 30$.
It is caused by the mean-field energy of boson-fermion repulsive interaction.
At sufficiently strong boson-fermion interaction strength,  however, the effect of the chemical potential $\mu_{\rm F}^r$ is significant and the total energy turns to decrease.
We plot the behavior of the chemical potential of the superfluid states $\mu_{\rm F}^r$ as a function of $U_{\rm BF}/U_{\rm BB}$ in Figure~\ref{fig2}(b).
In the weak boson-fermion coupling regime, the relation $\mu_{\rm F}^r \simeq \epsilon_{\rm F}$ holds as in the ordinary BCS theory.
However in the strong boson-fermion coupling regime, $\mu_{\rm F}^r$ starts to decrease, eventually becoming negative with increasing $U_{\rm BF}/U_{\rm BB}$.
Negative values of the chemical potential indicate formation of bound pairs~\cite{Leggett,Noz-Schmi}.
%
%

\begin{figure}
\begin{center}
      \includegraphics[width=9.cm]{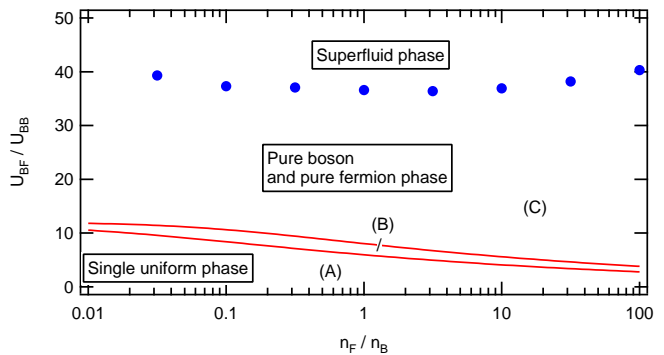}
\end{center}
\caption{
Phase diagram with considering the superfluid transition at $n_{\rm B} = 10^{14}  \rm cm^{-3}$.
Circles denote the boundary between the \textit{p}-wave superfluid phase with axial state and normal phase with phase separated state.
}
\label{fig3}
\end{figure}

The dot-dashed line in Fig.~\ref{fig2}(a) corresponds to total energy per volume of the ground state in normal phase, which is the same as the result in Fig.~\ref{fig1}(b).
The direct comparison between the ground states in the two phases shows that phase transition
from the phase-separated state in normal phase to the axial state in superfluid phase occurs
when the interaction ratio $U_{\rm BF}/U_{\rm BB}$ increases.
We see that the energy of the axial state is lower than that of the polar state
in the strong boson-fermion coupling regime.
It is caused by the angle dependence of the energy gap.
The axial state has anisotropic energy gap in momentum space with zero energy gap on the north and south poles.
For the polar state, on the other hand, the energy gap vanishes on the equator in momentum space.
So the former state contributes to the decrease of the total energy more than the latter case.
Figure~\ref{fig3} shows phase diagram of the system in the $U_{\rm BF}/U_{\rm BB}$-$n_{\rm F}/n_{\rm B}$ plane for
$n_{\rm B}=10^{14}  {\rm cm^{-3}}$.
The lower and upper lines are boundaries of the phases (A)-(B) and the phases (B)-(C) in normal phase
as already explained in Fig.~\ref{fig1}(a).
The circle corresponds to the critical points below and above which the phase-separated state and axial
state realizes, respectively. The dependence on $n_{\rm F}/n_{\rm B}$ of the phase boundary is not strong.
Thus we conclude that $p$-wave superfluidity can be observed at sufficiently strong boson-fermion interaction compared to boson-boson interaction. This is the main result in the present paper.
Using typical experimental parameters, we estimate the \textit{s}-wave scattering length between boson and fermion simply from the rate of coupling constant.
It is estimated as $a_{\rm BF}/a_{\rm BB} \simeq 23$ that is tunable by the technique
of Feshbach resonance~\cite{RbK-tune,RbK-cont}.
%
%
%
%
%
%
%
%
%
%
%
%
\begin{figure}
\begin{center}
\includegraphics[width=8.cm]{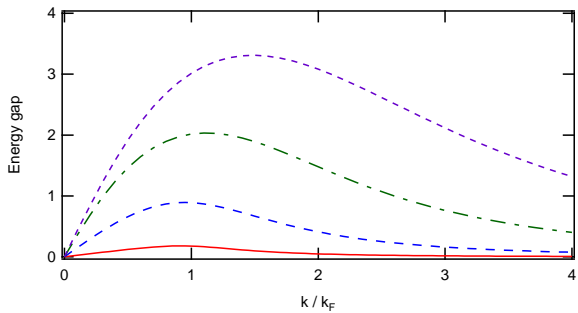}
\end{center}
\caption{
Momentum $|\mathbf k|$ dependence of the radial gap of the axial state $\Delta^{\rm ax}(k)$ scaled by $\epsilon_{\rm F}$ at $n_{\rm F} = n_{\rm B} = 10^{14} \rm cm^{-3}$.
Solid, dashed, dashed-dotted, and dotted lines denote results for $U_{\rm BF}/U_{\rm BB} = 20$, $30$, $40$ and $50$, respectively.
}
\label{fig4}
\end{figure}
%
\begin{figure}
\begin{center}
      \includegraphics[width=8.cm]{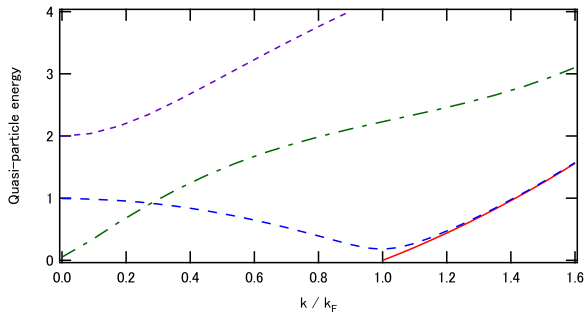}
\end{center}
\caption{
Quasi-particle energy in the axial state $E_{\bf k}$ scaled by $\epsilon_{\rm F}$ as a function of $k/k_{\rm F}$ with the angular component $\theta = \pi/2$ at $n_{\rm F} = n_{\rm B} = 10^{14} \rm cm^{-3}$.
Dashed, dashed-dotted, and dotted lines denote $U_{\rm BF}/U_{\rm BB} = 20$, $40$, and $50$.
And a solid line denotes dispersion relation of ideal fermi gases.
}
\label{fig5}
\end{figure}
%
%
%
\begin{figure}
\begin{center}
      \includegraphics[width=8.cm]{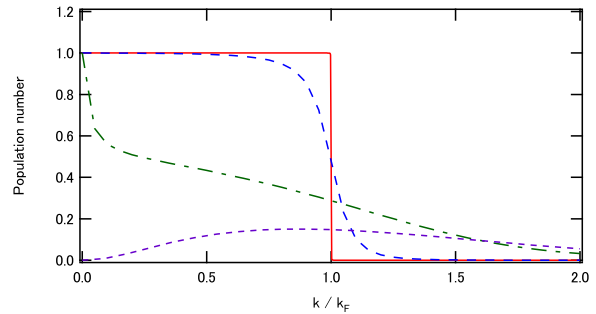}
\end{center}
\caption{
Population numbers of fermions, $v_{\mathbf k}^2 = ( 1 - \xi_{\mathbf k} / E_{\mathbf k} )/2$, in momentum space.
Solid, dashed, dashed-dotted, and dotted lines denote $U_{\rm BF}/U_{\rm BB} = 10$, $20$, $40$ and $50$, respectively.
}
\label{fig6}
\end{figure}
%
\subsection{$p$-wave pairing state in strong boson-fermion coupling regime}
We have just seen that $p$-wave superfluid state arises at sufficiently strong boson-fermion interaction strength.
Properties of the strong boson-fermion coupling superfluid state such as energy gap, quasi-particle, and momentum distribution  are expected to differ from those of the weak boson-fermion coupling one.
In the following we will show some aspects of the $p$-wave pairing in the strong boson-fermion coupling regime.

Figure~\ref{fig4} shows the radial component of energy gap of the axial state $\Delta^{\rm ax}(k)$ in unit of Fermi energy $\epsilon_{\rm F}$ as a function of $k/k_{\rm F}$ where $k_{\rm F} = (6\pi^2 n_{\rm F})^{1/3}$.
The energy gap peaks at around the Fermi surface, and a Cooper pair can be formed even in high momentum region of $|\mathbf k|>k_{\rm F}$.
The gap also becomes larger, and the width around the peak gets broader with increasing $U_{BF}/U_{BB}$.
The gap vanishes at $k=0$, and it originates from the property of the interaction with angular momentum $l=1$ between fermions.
%
%
%
%
%

Since $\Delta(\bf k)$ corresponds to a pairing potential that affects on the pairing state of ($\bf k$,  $-\bf k$),
the gap crucially influences quasi-particle spectrum, see Figure~\ref{fig5}.
The quasi-particle energy scaled by Fermi energy $\epsilon_{\rm F}$ is plotted as a function of $k/k_{\rm F}$.
The solid line denotes dispersion relation of a free fermion.
In the weak boson-fermion coupling regime, the quasi-particle energy differs from that of a free fermion only around the Fermi surface.
In the strong boson-fermion coupling regime, the quasi-particle spectrum changes dramatically.
For $\mu_{\rm F}^r < 0$, the gap opens at zero momentum and corresponds to molecular binding energy~\cite{Leggett,Noz-Schmi}.
We can see that a Cooper pair is now a tightly bound state of $l=1$ like a diatomic molecule with the binding energy $-2|\mu_{\rm F}^r|$ in the strong boson-fermion coupling regime.
%
%
%
%
%
%
%

Finally, the momentum distribution function of fermions for different $U_{\rm BF}/U_{\rm BB}$ values is shown in Figure~\ref{fig6}.
As $U_{\rm BF}/U_{\rm BB}$ increases, the occupation of fermions reduces significantly and
Fermi statistics turns to be less important.
The momentum distribution at large $U_{\rm BF}/U_{\rm BB}$ is expected to be proportional to
momentum distribution of $l=1$ bound state of two fermions.
The behavior of the distribution at $k = 0$ is explained as follows.
From Eq.(\ref{Nconservation}) the population number of fermions at $k = 0$ becomes $v_{\bf k=0}^2 = (1+\mu_{\rm F}^r/|\mu_{\rm F}^r|)/2$.
In the $U_{\rm BF}/U_{\rm BB}=10$, $20$, and $40$ cases, the population number of fermions at $k = 0$ becomes $v_{\bf k=0}^2 = 1$ because of $\mu_{\rm F}^r > 0$ shown in Fig.~\ref{fig2}(b).
In the $U_{\rm BF}/U_{\rm BB}=50$ case with $\mu_{\rm F}^r < 0$, however, the population number of fermions vanishes at $k = 0$.
This result can be also understood by noting the following fact.
For $\mu_{\rm F}^r > 0$, as indicated by Fig.~\ref{fig4}, fermions of extremely small momentum
are almost non-interacting and occupy as free particles.
For $\mu_{\rm F}^r < 0$, the population number distribution tends to have the form of
momentum distribution of single bound pair wave function~\cite{Noz-Schmi}
and  vanishes at $k=0$ because of $l = 1$ bound state.
The particle distribution is observable in the real experiments, for example,  by using the time-of-flight method.
Thus we consider this result as an indication for the observation of the $p$-wave superfluid.
%
%
%
%
%
%
\section{Summary}
We considered a system of repulsively interacting Bose-Fermi mixtures of spin polarized uniform
atomic gases at zero temperature.
We investigated the possibility of realization of $p$-wave superfluidity of fermions due to effective 
attractive interaction via density fluctuations of BEC in the case of strong boson-fermion interaction within mean-field approximation.
By direct energy comparison between $p$-wave superfluid and phase-separated states, 
we found that $p$-wave superfluidity can be observed at sufficiently strong boson-fermion interaction compared to boson-boson interaction.

We also discussed unique features of the ground state in strong coupling $p$-wave superfluidity.
We calculated the quasi-particle energy and found that a Cooper pair is a tightly bound state like a diatomic molecule in the strong boson-fermion coupling regime.
We also calculated the momentum distribution function of fermions.
In the strong boson-fermion coupling regime, we showed the property of the single bound pair wave function and suggested an observable indication of $p$-wave superfluid in the real experiment.

The present analysis was made by using the standard mean-field approach with BCS wave function~\cite{Leggett}.
The results at very strong coupling would be modified by fermionic self-energy correction as well as by dynamical screening of the effective interaction~\cite{StrongBF}.
The consideration of these effects is left for a future study.
Nevertheless, the mean-field approaches are expected to provide good ways of building up intuition.
%
%
%
%


\begin{references}
\bibitem{note1} We do not expect a superfluid state in the phase-separated system, as the phonon-mediated attraction between fermions is reduced in this case.
\end{references}
\end{document}